\documentclass[a4paper,11pt]{article}
\usepackage{pos}

\newcommand{\snn}{$\sqrt{s_{\mathrm{NN}}}~$}
\newcommand{\kst  }{$\mathrm{K^{*0}}~$}
\newcommand{\ksta  }{$\mathrm{K^{*0}}$}
\newcommand{\ks}{$\mathrm{K^{0}_{S}}~$}
\newcommand{\ksa}{$\mathrm{K^{0}_{S}}$}
\newcommand{\pt}{$p_{\mathrm{T}}$}
\newcommand{\ph} {$\mathrm{\phi}~$}
\newcommand{\pha} {$\mathrm{\phi}$}
\newcommand{\rh}{\ensuremath{\mathrm{\rho_{00}}}}
\newcommand{\s}{$\sqrt{s}~$}

\title{Spin alignment measurements of vector mesons with ALICE at LHC}
\ShortTitle{Spin alignment measurements}

\author*[a]{Bedangadas Mohanty (for the ALICE Collaboration)}

\affiliation[a]{National Institute of Science Education and Research, HBNI, Jatni 752050, India}

\emailAdd{bedanga@niser.ac.in}

\abstract{We present the measurements related to the spin alignment of
  \kst and \ph vector mesons at mid-rapidity for Pb--Pb collisions at
  \snn = 2.76 TeV using the ALICE detector at the LHC. The second diagonal spin density matrix
element (\rh) is measured from the angular distribution of the decay
daughters of the vector meson in the decay rest frame, with respect to the 
 event plane and the production plane. The \rh~values are found to be less
 than 1/3 at low \pt~($<$ 2 GeV/$c$) for both vector mesons. The
 observed deviations from 1/3 are maximal for 
 mid-central collisions at a level of 3$\sigma$ for \kst and 2$\sigma$
 for \ph mesons. As a control measurement the \rh~values were found to
 be consistent with 1/3, i.e., no significant spin alignment, for pp
 collisions with respect to the production plane. The results are
 qualitatively consistent with the expectation from a model that
 attributes the spin alignment to a polarization of quarks in the presence of
 large initial angular momentum and a subsequent hadronization by the
 process of  recombination.} 

\FullConference{%
  40th International Conference on High Energy physics - ICHEP2020\\
  July 28 - August 6, 2020\\
  Prague, Czech Republic (virtual meeting)
}

\begin{document}
\maketitle

\section{Introduction}
\label{sec:intro}
The initial stages of high energy heavy-ion collisions at non-zero
impact parameter have been theoretically shown to have a large magnetic
field~\cite{Kharzeev:2007jp} and angular
momentum~\cite{Becattini:2007sr}. Specifically, the effect of angular
momentum, a conserved quantity, is expected to be felt throughout the evolution of the
system formed in the collision. One of the physics 
interests of the heavy-ion program using the ALICE detector at the LHC is to look for signatures of the 
spin-orbit interactions. This can be studied by measuring the angular distributions 
of the decay daughters of hyperons and vector mesons relative to the system's
angular momentum direction~\cite{Abelev:2007zk,STAR:2017ckg,Abelev:2008ag}.

The angular distributions are measured with respect to a quantization axis, which can either be perpendicular 
to the production plane of the vector meson, or normal to the reaction
plane of the system. The normal direction is expected to be the
direction of the angular momentum for the system. The production plane is 
defined by the momentum of the vector meson under study and the beam direction, whereas the reaction plane is 
defined by the impact parameter and beam direction. The spin alignment of a vector meson is described by a 3 $\times$ 3 
Hermitian spin-density matrix~\cite{Yang:2017sdk}.  The matrix
elements can be obtained by measuring the angular 
distributions of the decay products of the vector mesons with respect 
to a quantization axis. The angle denoted as $\theta^{*}$ is that 
defined by one of the decay daughters of the vector meson in the rest
frame of the vector meson with respect to the quantization 
axis. In general, the angular  distribution for vector mesons is expressed as ~\cite{Schilling:1969um}
\begin{equation}
\frac{\mathrm{d}N}{\mathrm{d}\cos{\theta^{*}}} = N_{0} [ 1 - \rho_{00} + \cos^{2}{\theta^{*}}(3\rho_{00} - 1 ) ],
\label{eqn1}
\end{equation}
$N_{0}$ is a normalization constant, $\rho_{00}$ is the second
diagonal element of the spin density matrix. 
The spin 1 vector mesons can be in three spin
states of --1, 0 and 1.
The \rh~defines the probability of finding a vector meson in spin
state zero. The \rh~is~1/3 in the absence of spin alignment and the angular
distribution in Eq.~\ref{eqn1} is uniform.  If initial conditions or the final hadronization process cause polarization effects in heavy-ion collisions, 
then the angular distributions as defined in Eq.~\ref{eqn1}  would become non uniform. 
This would lead to $\rho_{00}$ values being different from 1/3. This is the experimental signature of spin alignment.

In this contribution, we present the results at LHC energies related to the
spin alignment of  \kst  and \ph  vector 
mesons through the measurement of $\rho_{00}$ in pp and Pb--Pb collisions with respect to the production 
plane (PP) and event plane (EP, an experimental measure of the
reaction plane).

\section{Analysis details}

The analyses are carried out using 14 million minimum bias Pb--Pb
collisions at \snn = 2.76 TeV, collected in the year 2010 and  43
million minimum bias pp collisions at $\sqrt{s}$ = 13 TeV, taken in
the year 2015.  The measurements for vector mesons are performed at
midrapidity ($|y| <$ 0.5) as a function of the transverse momentum \pt~and for different
centrality classes in Pb--Pb collisions. The \ks analysis is performed
only for Pb--Pb collisions in the 20--40\% centrality class as a null
hypothesis test, as it has spin zero. The details of the ALICE
detector, trigger conditions, centrality selection, and second order
event plane estimation using the V0 detectors at
forward rapidity, can be found
in~\cite{Abelev:2014pua, Aamodt:2008zz,Aamodt:2010cz,Abelev:2013qoq}.

The \kst and \ph vector mesons are reconstructed via their decays into
charged K$\pi$ and KK pairs, respectively, while the \ks is
reconstructed via its decay into two pions. The Time Projection
Chamber (TPC)~\cite{Alme:2010ke} and Time-of-Flight (TOF)
detector~\cite{Dellacasa:2000kh} are used to identify the decay
products of these mesons via specific ionization energy loss and
time-of-flight measurements, respectively.  
The meson yields are determined via the invariant mass
technique~\cite{Adam:2017zbf,Abelev:2014uua,Acharya:2019qge, Abelev:2013xaa}. The
background coming from combinatorial pairs and misidentified particles
is removed by constructing the invariant mass distribution from the
mixed events~\cite{Adam:2017zbf,Abelev:2014uua,Acharya:2019qge, Abelev:2013xaa}.

The invariant mass distributions are fitted with a Breit-Wigner
(Voigtian: convolution of Breit-Wigner and Gaussian distributions)  
function for the \ksta(\pha) signal and a second order
polynomial that describes the residual background,  in order to
extract the yields~\cite{Adam:2017zbf,Abelev:2014uua,Acharya:2019qge}.  Extracted
yields are then corrected for the reconstruction efficiency and
acceptance in each $\cos{\theta^{*}}$ and \pt~
bin~\cite{Adam:2017zbf,Abelev:2014uua,Acharya:2019qge}.  
The resulting efficiency and acceptance corrected
$\mathrm{d}N/\mathrm{d}\cos{\theta^{*}}$ distributions are
fitted with the functional form given in Eq.~\ref{eqn1} to determine
$\rho_{00}$ for each \pt~bin in pp and Pb--Pb collisions~\cite{Acharya:2019vpe}. For the EP
results, the finite resolution of the event plane is corrected~\cite{Abelev:2014pua}. 

There are three main sources of systematic uncertainties in the
measurements of the angular distribution of vector meson decays, which
are propagated to \rh{}: (a) Meson yield extraction procedure. These
sources contribute to the uncertainties on the \rh~value at a level of
12(8)\% at the lowest \pt~and decrease with \pt~to 4(3)\% at the
highest \pt~studied for the \ksta(\pha). (b) Track selection
criteria. The systematic uncertainties on the \rh~value due to
variation on the track selection criteria are 14(6)\% at the lowest
\pt~and about 11(5)\% at the highest \pt~for \ksta(\pha). (c) Particle
identification procedure. The corresponding uncertainty is 5(3)\% at the lowest
\pt~and about 4(4.5)\% at the highest \pt~studied for \ksta(\pha).
The total systematic uncertainty on \rh{} is obtained by adding all
the contributions in quadrature and details can be found in
Ref.~\cite{Acharya:2019vpe}.

\section{Results and discussions}

\begin{figure}
      \begin{center}
        \includegraphics[scale=0.5]{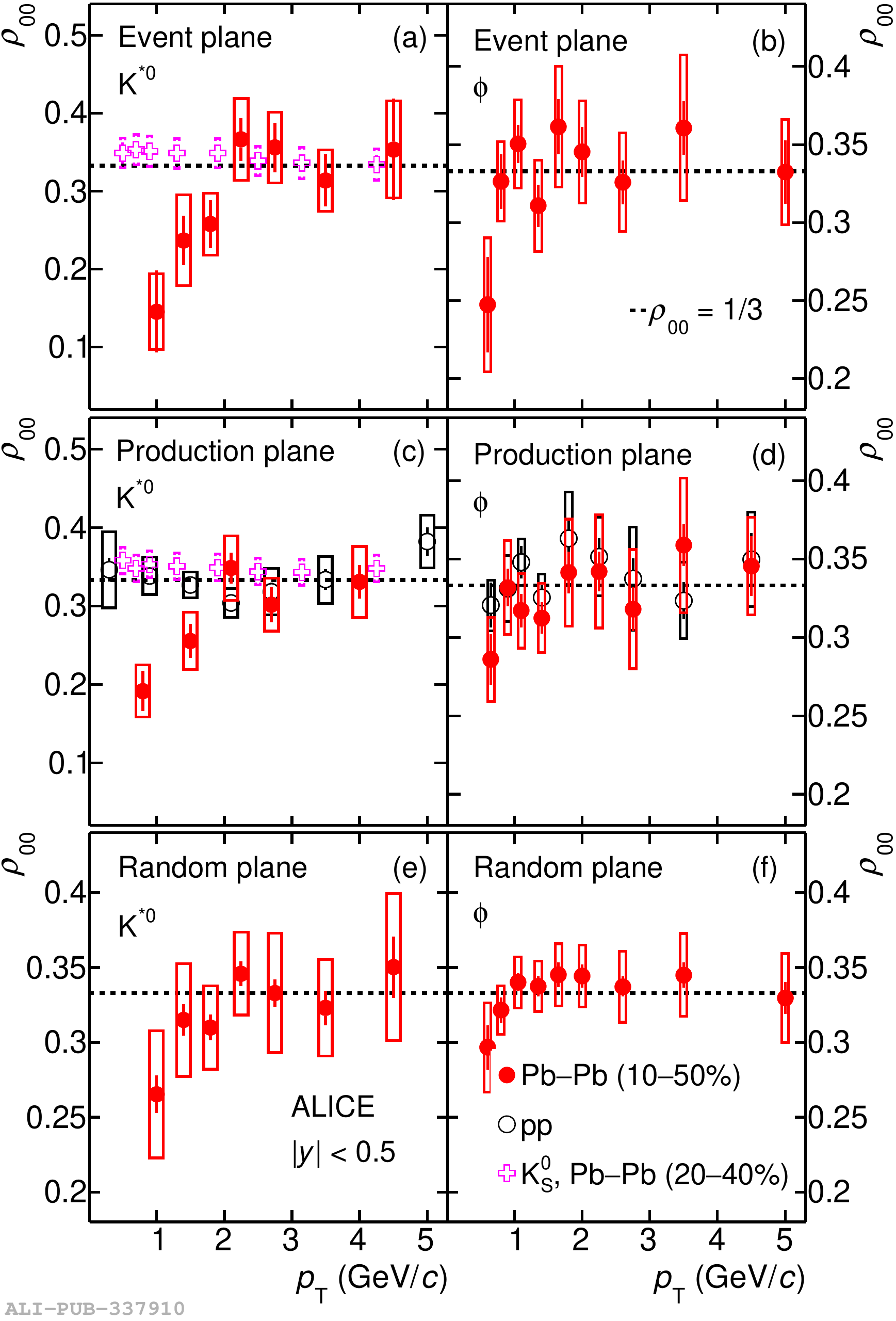}
       \caption{The measured \rh{} as a function of \pt~
       corresponding to \ksta, \ph, and \ks mesons at $|y| <$ 0.5 in Pb--Pb collisions at \snn = 2.76 TeV and minimum bias pp collisions at \s = 13 TeV.
       Results are shown for spin alignment with respect to event
       plane (panels a,b), production plane (c,d) and random event
       plane (e,f) for \kst (left column) and \ph (right column)~\cite{Acharya:2019vpe}.  The statistical and systematic uncertainties are 
          shown as bars and boxes, respectively.}
        \label{Fig:momentum}
      \end{center}
    \end{figure}

Figure~\ref{Fig:momentum} shows the measured \rh{} as a function of \pt~ 
for \kst and \ph mesons in pp and Pb--Pb collisions, along
with the measurements for \ks in Pb--Pb collisions. 
In mid-central (10--50\%) Pb--Pb collisions, \rh{} is below 1/3 at the
lowest measured \pt~and increases to 1/3 within uncertainties
for \pt~$>$ 2 GeV/$c$. At low \pt, the central value of \rh ~is
smaller for \kst than for \pha, although the results are compatible
within uncertainties. In pp collisions, \rh{} is independent of \pt~and
equal to 1/3 within uncertainties. For the spin zero hadron
\ksa, \rh{} is consistent with 1/3 within uncertainties in Pb--Pb collisions. 
The results with random event plane directions are also compatible
with no spin alignment for the studied \pt~range. The results for the random production plane (the momentum vector direction
of each vector meson is randomized) are similar to random event plane measurements.

    \begin{figure}
      \begin{center}
        \includegraphics[scale=0.5]{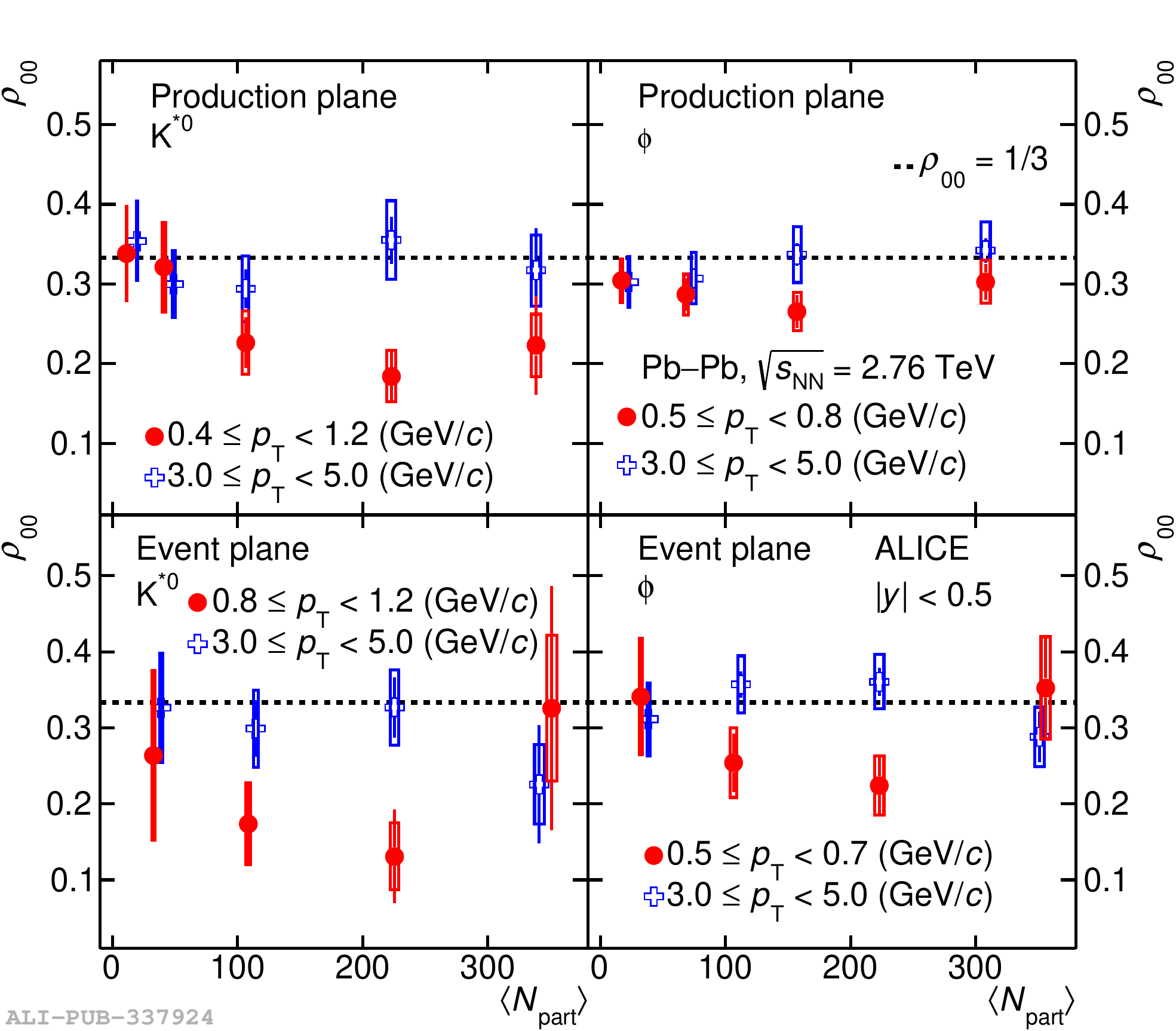}
       \caption{The measured \rh~as a function of $\langle
         N_{\mathrm {part}} \rangle$ for \kst and \ph mesons at low
         and high \pt~in Pb--Pb collisions~\cite{Acharya:2019vpe}. The statistical and systematic uncertainties are 
          shown as bars and boxes, respectively.}
        \label{Fig:Centrality}
      \end{center}
    \end{figure}

Figure~\ref{Fig:Centrality} shows \rh{} for \kst and \ph
mesons as a function of average number of participating
nucleons ($\langle N_{\mathrm {part}} \rangle$)~\cite{Abelev:2013qoq}  for Pb--Pb
collisions at \snn = 2.76~TeV. Large $\langle
N_{\mathrm {part}} \rangle$ correspond to the central collisions,
while peripheral events have low $\langle N_{\mathrm {part}} \rangle$.
In the lowest \pt{} range, the \rh~values have maximum  deviation from
1/3 for intermediate centrality (20--40\%) and approach 1/3 for both central and
peripheral collisions. 
At higher \pt, the \rh~measurements are consistent with 1/3 for all the collision centrality
classes studied for both vector mesons. 
For the low-\pt~measurements in mid-central Pb--Pb collisions, the maximum deviations of \rh~from 1/3 are 3.2 (2.6) 
$\sigma$ and 2.1 (1.9) $\sigma$ for \kst and \ph mesons, respectively, for
mid-central (20--40\%) Pb--Pb collisions with respect to the PP (EP). The $\sigma$ are
calculated by adding statistical and systematic uncertainties in
quadrature.

The qualitative theory predictions for the spin alignment
effect are the following~\cite{Liang:2004xn}: (i) \rh~$>$ 1/3 if the hadronization of a polarized parton
proceeds via fragmentation, (ii) \rh~$<$ 1/3 for hadronization of
polarized parton via recombination, (iii) \rh~is expected to have a maximum deviation from 1/3 for
mid-central heavy-ion collisions, whic is consistent with angular momentum being 
maximal, and a smaller deviation for both peripheral (large impact
parameter and small $\langle N_{\mathrm {part}} \rangle$) and
central (small impact parameter and large $\langle N_{\mathrm {part}} \rangle$) collisions, (iv)
the \rh~value is expected to have a largest deviation from 1/3 at low \pt~and
reach the value of 1/3 at high \pt~in the recombination hadronization scenario, and (v) the effect is
expected to be larger for \kst compared to \ph due
to their constituent quark composition. All of these features are
probed experimentally 
~for \kst and \ph vector mesons in Pb--Pb collisions~\cite{Acharya:2019vpe} and reported in
this contribution.
The experimental results indicate that a spin alignment is present at lower \pt,
which is qualitatively consistent with the  predictions~\cite{Liang:2004xn}. 
This centrality  dependence  is qualitatively consistent with the
dependence of initial angular momentum on  impact parameter in
heavy-ion collisions~\cite{Becattini:2007sr}.  

Significant polarization of $\Lambda$ baryons (spin = 1/2) was reported at
low RHIC energies. The polarization is found to decrease with increasing \snn~\cite{STAR:2017ckg}. At the LHC energies,
the global polarization for $\Lambda$ baryons was measured to be
compatible with zero within uncertainties~\cite{Acharya:2019ryw}. 
In the recombination model, \rh~is expected to depend on the square of
the quark polarization whereas the $\Lambda$ polarization depends
linearly on it, therefore using quark polarization information from
$\Lambda$ measurements will yield a \rh~$\sim$~1/3 at LHC
energies. The large effect observed for the central value of \rh~for
mid-central Pb--Pb collisions at low \pt~is therefore puzzling.
However, the magnitude of the spin alignment also
depends on the details of the transfer of the quark
polarization to the hadrons (baryon vs. meson), details of the
hadronization mechanism (recombination vs. fragmentation), re-scattering, regeneration, and
possibly the lifetime and mass of the hadrons in the system. Moreover, 
the vector mesons are predominantly primordially produced whereas the
hyperons are expected to have large contributions from resonance decays. 

\section{Summary}
We observe a significant spin alignment effect for \kst and \ph mesons in
heavy-ion collisions. The measured spin alignment of the vector mesons
is surprisingly large
compared to the polarization measured for $\Lambda$ hyperons.
The effect is strongest when the alignment is measured at low \pt~with
respect to a vector perpendicular to the reaction plane and  for mid-central (10--50\%) collisions. 
These observations are qualitatively consistent with expectations from
the effect of large initial angular momentum in non-central heavy-ion
collisions, which leads to quark polarization via spin-orbit coupling
and is subsequently transferred to hadronic degrees of freedom by
hadronization via recombination. Results with increased statistical
precision are expected for spin alignment studies with the Pb--Pb data
set at \snn = 5.02 TeV.  This will help in arriving at a better
significance of the observed effects. 
In future measurements, the
difference in the polarization of $\mathrm{K^{*\pm}}~$ 
and \ksta, due to their different magnetic moments, would be
directly sensitive to the effect of the large initial magnetic field produced in heavy-ion collisions.

\end{document}